# Power law in website ratings

*D.V. Lande*[1,2] *(dwl@visti.net), A.A. Snarskii*[2] *(asnarskii@gmail.com)*

[1]Elvisti, Kiev, Ukraine
[2]Kiev Polytechnical Institute, Kiev, Ukraine

In the practical work of websites popularization, analysis of their efficiency and downloading it is of key importance to take into account web-ratings data. The main indicators of website traffic include the number of unique hosts from which the analyzed website was addressed and the number of granted web pages (hits) per unit time (for example, day, month or year). As one of the hundreds of rating services one can cite a resource located at the address http://top.ucoz.com/. Fig. 1 gives an example of a page of this web rating on the subject "Business and Finances".

*Fig. 1. Fragment of rating service page http://top.ucoz.com/*

It is known that the major part of digital sequences characterizing the products of human activity obeys power law [1]. For example, in economics the variant of power distribution is known as the "Pareto rule" [2] (the distribution of social wealth, in particular, arable lands, among people), in linguistics the distribution of words in the text obeys Zipf's law [3] which is also described by power law. Zipf also showed that the distribution of cities according to the number of citizens obeys power law [4]. The same law is known to govern the distribution of the degrees of nodes in modern information networks (in particular, the Internet) [5].

Therefore, the authors have expressed and verified by numerous experiments the hypothesis of power distribution of the number of hosts $H$, as well as the number of hits $S$ depending on site ratings $r$ (numbers in rating):

$$H = r^{\alpha} C_h, \qquad (1)$$
$$S = r^{\beta} C_s, \qquad (2)$$

where $\alpha$, $\beta$ and $C_h$, $C_s$ are coefficients that in these equations are considered as constants.

Fig. 2 represents the plots of distribution of the number of hits per day depending on the position in rating for two knowledge domains – "Business and Finances" and "Games". Fig. 3 represents the plots of distribution of the number of unique hosts depending on the position in rating within the same period for the same knowledge domains.

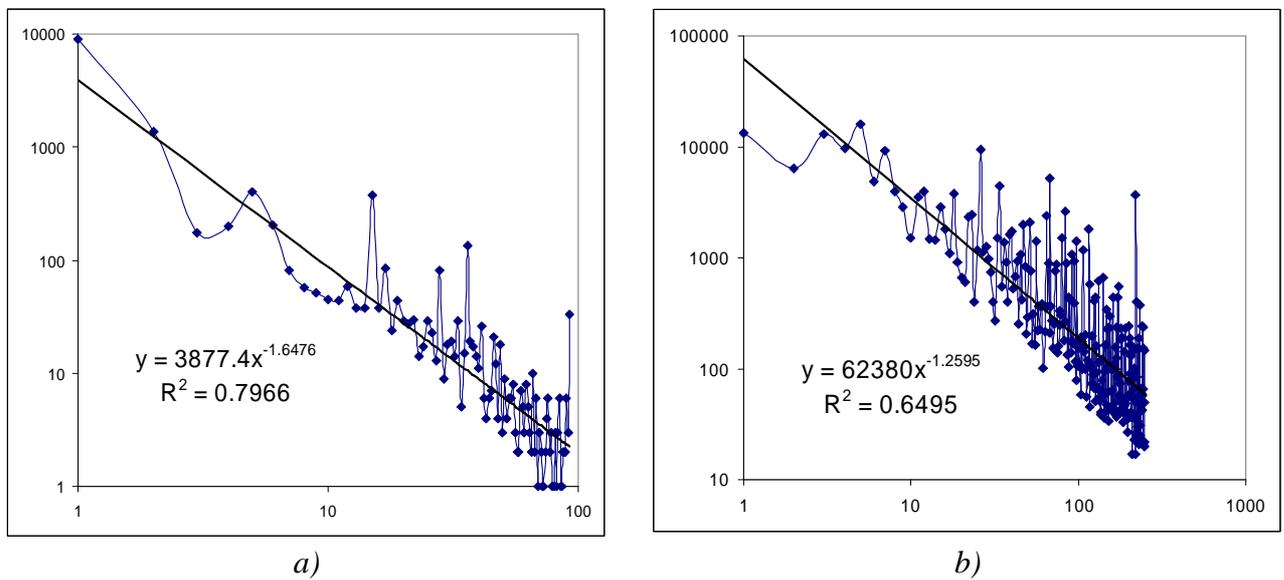

*Fig. 2. The plots of distribution of the number of hits (in the log-log scale):*
*a) – "Business and Finances" division; b) – "Games" division*

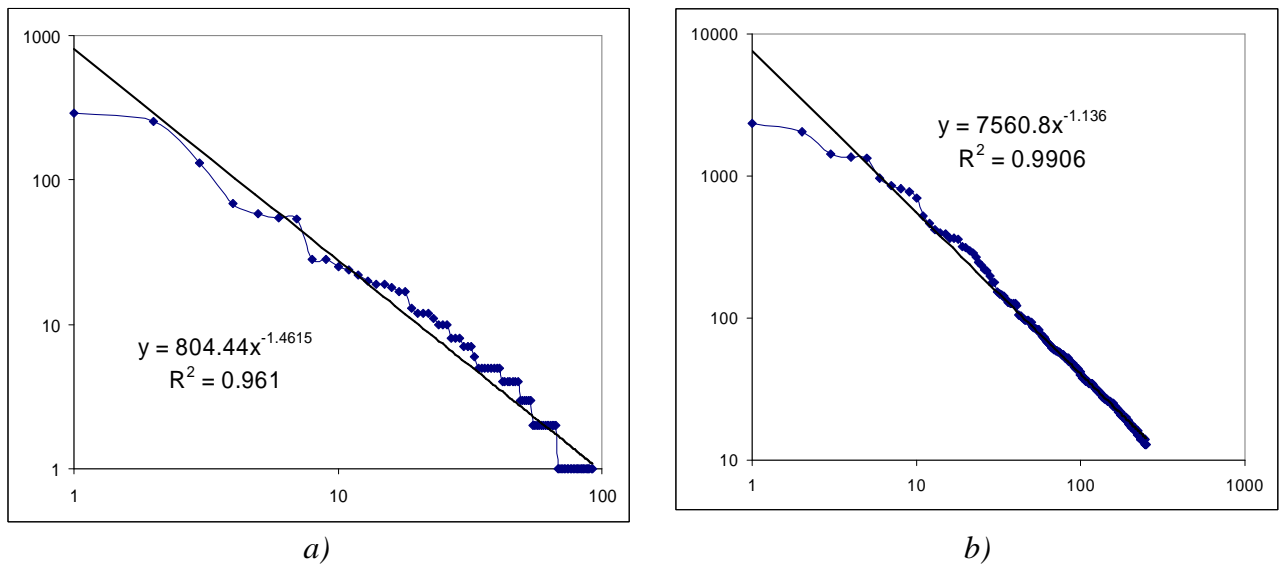

*Fig. 3. The plots of distribution of the number of unique hosts (in the log-log scale): a) –"Business and Finances" division; b) –"Games" division*

The plots represented in Fig. 2 and 3 (as well as numerous similar experiments conducted by the authors) give grounds to conclude that the distributions of hits and unique hosts obey power law, i.e. Zipf's law is valid for these cases as well. In so doing, a spread in values in case of unique hosts is considerably smaller than in case of hits. Apart from everything else, it is explained by the fact that the analyzed rating was composed according to the values of unique hosts. If rating of sites is constructed according to hits, then based on the same data, but classified in decreasing order of hit values, one can obtain a "more smooth" plot (as shown for the rating division "Games" in Fig. 4).

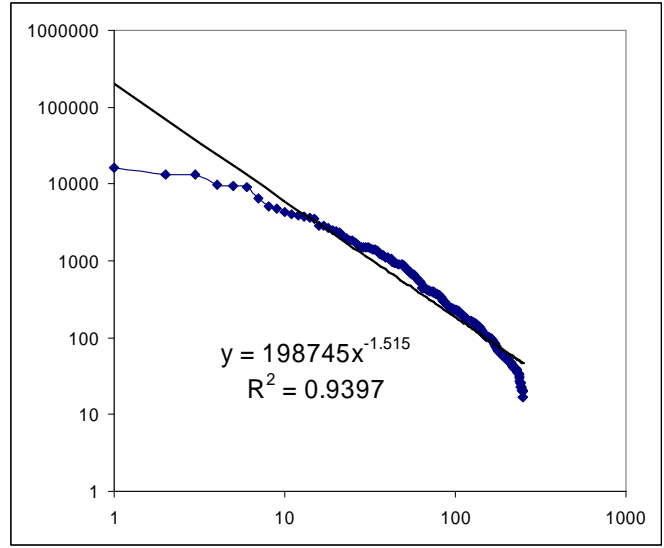

*Fig. 4. The plots of distribution of the number of hits (in the log-log scale) in case of rating according to hits*

If the number of unique hosts to a greater extent characterizes position of websites as the nodes in hypertext network, the number of hits to the largest degree characterizes the loading of websites and the interest of target groups of users in their content. Really, if the site is not interesting, the user having read a page of this site which he could reach, for example, from the search system, will no longer move by internal hyperlinks, but will quit this site. On the contrary, if the site is interesting, the number of internal jumps (hence, the hits) will be the largest.

Of certain interest is the ratio between the number of hits ($S$) and hosts ($H$). In practice there is even used such a concept as "average number of viewed pages" ($S/H$), which on default supposes a linear dependence of $S$ on $H$. What actually happens is that linear dependence is observed only as a partial case of power dependence, and not always. To obtain the necessary dependence, we will define the value of $r$ from (1):

$$r = H^{1/\alpha} / C_h^{1/\alpha},$$

following which, we will substitute the resulting expression to (2):

$$S = H^{\beta/\alpha} C_s C_h^{-\beta/\alpha} = H^{\gamma} C_{s/h}, \qquad (3)$$

where $C_{s/h}$ and $\gamma$ are also constants determined by the values $C_s C_h^{-\beta/\alpha}$ and $\beta/\alpha$, respectively.

Conducted experiments (see Fig. 5) prove the obtained theoretical result.

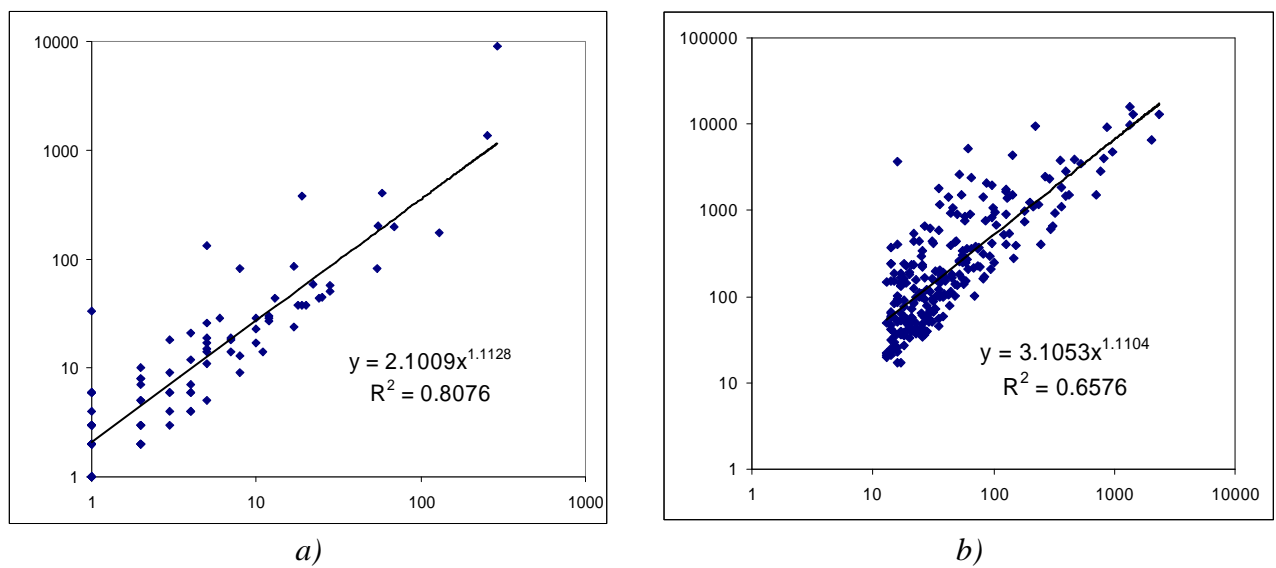

*Fig. 5. The plots of dependence of the number of hits on the number of unique hosts per day (in the log-log scale): a) – "Business and Finances" division; b) –"Games" division*

**Conclusions**

1. Another new power law has been discovered on the Internet, in particular, on the WWW.
2. In some cases the value of $\gamma$ in (3) is so much close to unity that one can speak about a linear dependence.
3. The obtained relation can be used by the web analysts for:
    - analysis of sites, deviations of the values of real ratio between hits and hosts from the theoretical, typical of this knowledge domain;
    - prediction of web servers loading and calculation of their efficiency;
    - prediction of advertising campaigns on the websites.